\documentclass[5p,times,twocolumn]{elsarticle}
\biboptions{comma,sort&compress}
\usepackage[utf8]{inputenc}
\usepackage{bm}
\usepackage{hyperref} 
\usepackage[dvipsnames]{xcolor}
\usepackage{nicefrac}

\usepackage{amsmath,amsfonts,amsthm,amssymb}
\usepackage{nicefrac}
\usepackage{yfonts}
\usepackage{slashed}
\hypersetup{
colorlinks = true,
linkcolor = blue,
anchorcolor = blue,
citecolor = blue,
filecolor = blue,
urlcolor = blue
}
\usepackage{ulem}
\graphicspath{{fig/}}

\def\ps@pprintTitle{%
 \let\@oddhead\@empty
 \let\@evenhead\@empty
 \def\@oddfoot{}%
 \let\@evenfoot\@oddfoot}
\makeatother

\journal{Physics Letters B}

\begin{document}

\begin{frontmatter}



\title{Unravelling theoretical challenges in understanding $B_c$ meson decays}


\author[a]{Sonali Patnaik}
\ead{sonali\_patnaik@niser.ac.in}
\address[a]{School of Physical Sciences, National Institute of Science Education and Research, An OCC of Homi Bhabha National Institute, Jatni-752050, India.}

\begin{abstract}
The $B_c$ meson, a unique bound state comprising of two open heavy flavors, charm and bottom, offers a rich avenue for probing the predictions of the Next Decade - Standard Model (ND-SM) physics properties due to its heavy mass. With recent observations of its excited states, interest in understanding $B_c$ production mechanisms and decay modes has surged. This article presents the current state of art on $B_c$ mesons, encompassing production mechanisms, properties of different decay modes, and theoretical modeling. We present novel findings on the newly constructed ratios $(\mathcal{R}_{\eta_c/J/\psi}$, $\mathcal{R}_{D/D^*})$ in semileptonic and ($\mathcal{R}_\mu^\tau)^{B_c}$ , ($\mathcal{R}_\mu^\tau)^{B_c^*}$ in leptonic decays, respectively. These results emphasize the importance of $B_c$ studies in the future collider experiments. The article further explores CP effects in $B_c$ meson decays refining our understanding of heavy flavor properties. Finally, potential avenues for future research, and leveraging upcoming collider experiments are outlined.
\end{abstract}



\end{frontmatter}




\section{Introduction}
\label{sec:intro}
The $B_c$ meson, typically denoting the fundamental $1^1 S_0$ state of $(c \bar b)$-quarkonium with a combination of two distinct heavy flavors, presents a distinct bound-state arrangement ideal for scrutinizing the predictions of the Standard Model (SM). This feature has generated considerable attention from researchers because of its heavy mass. Ever since the discovery of $B_c$ meson in Fermilab by the Collider Detector (CDF) Collaborations \cite{CDF:1998ihx} in 1998, the experimental probe to detect its family members in their ground and excited states continues over last two decades. With the observation  of around 20 occurrences of $B_c$ events in the $B_c \to J/\psi l \nu_l$ decay mode, highlights the potential for experimental investigation of $B_c$ meson at hadron colliders\cite{Wester:2005ex,Cheung:1999ir}. A comprehensive investigation of the $B_c$ family members is anticipated at the Large Hadron Collider (LHC), where the increased energy and significantly higher luminosity will enable more detailed studies. The lifetime of $B_c$ has been measured \cite{D0:2008bqs,D0:2008thm,CDF:2006kbk,CDF:2007umr} using decay channels: ${B_c}^{\pm} \rightarrow J/\psi e^{\pm} \nu_e$ and ${B_c}^{\pm} \rightarrow J/\psi \pi^ {\pm}$.  A more precise measurement of $B_c$ lifetime: $\tau_{B_c}=0.51^{+0.18}_{-0.16}(stat.)\pm 0.03(syst.)$ ps and its mass: $M =6.40\pm 0.39\pm 0.13\; \text{GeV}$ have been obtained \cite{LHCb:2014ilr} using the decay mode $B_c \rightarrow J/\psi \mu \nu_\mu X$, where $X$ denotes any possible additional particle in the final state. 

Research into the production of the $B_c$ meson holds significance for understanding this particle, as it elucidates the quantity of $B_c$ events feasible in a collider setting. Conversely, once the production rate is determined, theoretical models regarding $B_c$ production can be validated through the measurement of its cascade decays. Indeed, their production yields are significantly smaller than those of the charmonium and bottomonium states. The production process for the $\bar b c$ system differs significantly from that of the $\bar b b$ system, primarily due to the necessity of creating two heavy quark-antiquark pairs during a collision. While the $\bar b b$ pair can be generated through parton processes like $\bar q q$, $gg \to b\bar b$ at the level of $\alpha_s^2$, the fundamental mechanism for creating the $\bar b c$ system is considerably more complex, at least proportional to the fourth power of the strong coupling constant, $\alpha_S^4$: $q\bar q$ $gg \to (\bar b c) b \bar c$. At energies typical of the Tevatron and LHC, the gluon-gluon contribution dominates since the production of two pairs of heavy quarks is required. With an LHC luminosity of approximately 
$\mathcal{L} = 10^{34} cm^{-2}s^{-1}$, an estimated yield of around $5 \times 10^{10}$ $B_c$ events per year could be anticipated \cite{Gouz:2002kk}.

At present, most studies of the $B_c$ production focus on the hadronic production due to the high production rate at a hadron collider, such as Tevatron or LHC \cite{Chang:1992jb,Chang:2004bh,Chang:1992pt,Chang:2005wd,Cheung:1999ir,Chen:2018obq}. Indeed, the $B_c$ meson has been only observed at hadron colliders up to now \cite{CDF:1998axz,CDF:1998ihx,CDF:2012ksy,LHCb:2012ihf,LHCb:2014ilr,LHCb:2017lpu,LHCb:2019bem,CMS:2019uhm}. In addition to hadron colliders, a high luminosity $e^+$ $e^-$ collider can be a potentially good platform for studying the $B_c$ meson. Compared with the hadron collider, there are less backgrounds at the $e^+$ $e^-$ collider, thus it is suitable for studying the $B_c$ meson precisely. Moreover, the production mechanism of the $B_c$ meson at the $e^+$ $e^-$ collider is much simpler than that at the hadron collider. The $B_c$ mesons cannot be produced at the B-factories, which operate at the $\Upsilon(5S)$ peak. The reason is that the $\Upsilon$ has sufficient mass to produce up to a pair of $B_s$ but not a pair of $B_c$ mesons. As a result, the production of $B_c$ mesons is exclusive to the LHC. Recent research on the production of $B_c$ mesons at $e^+e^-$ colliders is detailed in Refs~ \cite{Chen:2020dtu,Zheng:2018fqv}.

The $B_c$ meson provides a specific possibility for investigating the production and decay mechanisms of heavy mesons. The reason is two fold: first, it is only the $B_c$ that contains two heavy quarks and at the same time carries explicit flavor numbers since for the top quark there is not enough time to form hadrons before decaying weakly due to the large top quark mass \cite{D0:1995jca}, secondly, because of the absence of decays caused by strong and electromagnetic interactions, the lightest pseudoscalar state $B_c$ can only decay weakly and is therefore long lived. Moreover, the $B_c$ meson serves as a favorable context for determining the Cabibbo-Kobayashi-Maskawa (CKM) matrix elements $V_{cb}$ and $V_{ub}$. Precise calculations of hadronic parameters are crucial for extracting CKM matrix elements by comparing SM predictions with experimental data on exclusive flavor-changing decay rates.

Like heavy quarkonia such as $\eta_c$, $J/\psi$, $\eta_b$, and $\Upsilon$, among others, the production cross section of the $B_c$ meson can also be studied down using factorization theories such as non-relativistic QCD (NRQCD) \cite{Bodwin:1994jh}. This approach enables the calculation of significant details regarding the production of $B_c$ through perturbative methods. Heavy quarkonia mesons, composed primarily of heavy (i.e., $c$ or $b$) quark pairs in a valence approximation, serve as a unique laboratory for studying strong interactions across both perturbative and non-perturbative domains. On a partonic level, the processes involving their production and decay can be elucidated in terms of the creation or annihilation of heavy quarks, which subsequently hadronize into the observed particles. Doubly heavy mesons with an open flavor occupy an intermediate position between charmonium and bottomonium particles, allowing them to be utilized as a testing ground for models devised to characterize heavy quarkonia particles with hidden flavor. In contrast to quarkonium production, where the color-octet contribution typically holds significance \cite{Brambilla:2010cs, Andronic:2015wma, Lansberg:2019adr} the production mechanism of the $B_c$ meson is more straightforward. This is because the color-octet contribution is consistently suppressed in $B_c$ production processes compared to the color-singlet contribution. 






The observations of the 2S excited states, $B_c(2^{1}S_0)$ and $B_c^{*}(2^{3}S_1)$, have been reported by the CMS and LHCb collaborations\cite{CMS:2019uhm,LHCb:2019bem}. As the $B_c$ meson encompasses two distinct heavy flavors, its excited states below the $B \bar D$ threshold typically undergo direct or indirect decay to the ground state through electromagnetic or strong interactions with nearly 100\% probability. These excited states serve as significant sources of the $B_c$ meson. Thus, the study on the production of the two excited states is inherently intriguing. In the upcoming sections, we will analyze the various decay processes of the $B_c$ meson and highlight new findings on newly constructed ratios in semileptonic and leptonic decays in view of the development of advanced energetic colliders, probing more understanding in the Unexplored Physics (UP) sectors. 

\section{Different decay properties of \texorpdfstring{$B_c$}{} meson}
\label{sec_decay}

We expect three major contributions to the $B_c$ decay width:
\begin{itemize}
    \item $\bar b \to \bar c W^+$ with $c$ as a spectator, leading to final states like $J/\psi \pi$ or $J/\psi l \nu_l$ 
    \item $c \to s W^+$, with the $\bar b$ as spectator, leading to final states like $B_s$ $\pi$ or $B_s l \nu_l$
    \item $c \bar b \to  W^+$ annihilation, leading to final states like $D$, $K$, $\tau$ $\nu_{\tau}$ or multiple pions.
\end{itemize}
Since these processes lead to different final states, their amplitudes do not interfere. The (CKM) matrix elements are strongly in favor of the c quark decays, even though the phase space of the c quark decays is significantly smaller than that of the b quark decays (i.e., $|V_{cs}|\gg |V_{cb}|, |V_{cd}|\gg |V_{ub}| $) \cite{Gouz:2002kk}.

The detection of ground and excited charmonium states and measured observables in the non-leptonic and semileptonic $B_c$-decays to charmonium ground and excited states are of special interest as it is easier to identify them in experiments. Studying weak decays of heavy quarks is made easier by the fact that the $B_c$ meson can only decay through weak interactions because it is below the $B \bar D$ threshold. The $B_c$ meson decays differ somewhat from the $B$ or $B_s$ meson decays because of the two heavy quarks varying their decay rates. While $b \to c$ decays have a wider phase space than the $c \to s$ transition, the CKM matrix element $|V_{cs}| \sim 1$ is significantly greater than $|V_{cb}| \sim 0.04$. As a result, the dominant decay width of the $B_c$ meson (about 70\%) is contributed by the c-quark decays \cite{Gouz:2002kk}.

\subsection{Semi-leptonic decays}
In the context of Lepton Flavor Universality (LFU) tests, $B_c$ decays have not been extensively explored yet. However, $B_c \to J/\psi$ decays is linked to $R_{D^{(*)}}$ via the same $b \to c l \nu_l$ currents. The LHCb collaboration \cite{LHCb:2017vlu} has determined the ratio $R_{J/\psi}^{LHCb}$, based on data collected in 2016 and total integrated luminosity of 3 ${\rm fb}^{-1}$. The result exceeds by 2 $\sigma$ from the SM prediction of $R_{J/\psi} = 0.2582(38)$ \cite{Harrison:2020nrv}
\begin{equation}
 R_{J/\psi}^{\text{LHCb}} =
    \frac{\mathcal{B} (B_c \to J/\psi \mu \nu)}{\mathcal{B} (B_c \to J/\psi e \nu)} =  0.71 \pm 0.17 \pm 0.18
\end{equation}

Additionally, very recently, the CMS collaboration \cite{CMS24:2024CMS} released the initial preliminary outcome of $R_{J/\psi}$ using the B parking data, 
\begin{equation}
    R_{J/\psi}^{\text{CMS}} = 0.17^{+ 0.18}_{- 0.17\; \text{Stat.}} \,^{+ 0.21}_{\;- 0.22\; \text{Syst.}} \,^{+ 0.19}_{\;- 0.18\; \text{Theory}}
\end{equation}
This finding aligns with the SM prediction \cite{Harrison:2020nrv} within a margin of 1.3 $\sigma$. Despite the significant experimental uncertainties in these results, it would be beneficial in the future to examine various scenarios of UP to address the anomalies observed in $R_{D^{(*)}}$. To obtain the exclusive semileptonic widths and their ratio, one needs the $B_c \to J/\psi$ hadronic form factors in the whole region of momentum transfer squared, $0 \le q^2 \le (m_{B_c} - m_{J/\psi})^2$. Recently, these form factors have been calculated using lattice QCD \cite{Harrison:2020gvo,Harrison:2020nrv} with an appreciable accuracy, thereby challenging all previous model calculations based on continuum QCD. A specific property of the $B_c \to$ charmonium transitions is the fact that all valence quarks are heavy. This, generally, enables us to represent the $B_c \to J/\psi$ form factors in terms of an overlap of heavy quarkonia wave functions. The $B_c \rightarrow J/\psi (\eta_c)$ form factors are often far less known than the $B^0 \rightarrow D^{(*)}$ ones. The reason being, two heavy quark flavors have been shown to emerge in the initial $(b\bar c )$ and final $(c\bar c )$ states. In the infinite heavy quark limit, the latter reduces the number of form factors by breaking the heavy quark symmetry (HQS) and leaving the residual heavy quark spin symmetry (HQSS). However, the HQSS does not fix the normalization of the form factors as the HQS does, for example, in the case of $B^0 \rightarrow D^{(*)}$. 

While the lepton component can be easily estimated using ordinary methods, the main theoretical challenge in studying exclusive semileptonic decays of bottom mesons relates to the calculation of the form factors, which parameterize the hadronic matrix elements. Since differential distributions are highly sensitive to momentum transfer squared $q^2$, it is crucial to accurately ascertain these form factors' behavior throughout the whole accessible kinematical range. Given the wide $q^2$ range of heavy-to-heavy and light decays, this is particularly significant. The majority of theoretical methods identify these form factors within a restricted $q^2$ range or at specific kinematical moments, and then use model parametrizations to extrapolate them. As an illustration, light cone sum rules \cite{Ball:2004ye} compute form factors at the final meson's maximum recoil point, $q^2$ = 0 (ignoring the small electron mass), whereas lattice QCD calculations examine the high $q^2$ region (refer to, for example, Ref. \cite{Heger:2021gxt} and its references). Additionally, the heavy quark ${1}/{m_Q}$ expansion is used in the majority of analyses of heavy-to-heavy semileptonic decays.

Along these lines, a broad variety of approaches was applied in the past, from a non-relativistic quark model1 to non- relativistic QCD~\cite{Zhu:2017lqu,Qiao:2011yz,Patnaik:2017cbl,Patnaik:2018sym,Patnaik:2019jho,Patnaik:2022moy,Patnaik:2023efe,Patnaik:2023ins}.
The main challenge is due to the presence of neutrinos, that escape detection, in the final states of the decays involving the $R_{J/\psi}$ ratio, which results in the absence of a distinct mass peak for the $B_c$ meson. This aspect adds complexity to the analysis, and it makes the derivation and estimation of backgrounds very central. A correct background estimation is therefore crucial, and it requires accounting for all potential decays of interest. The background coming from the mis-identification of muons is especially a critical point of the analysis, due to its significant impact on the results.


The transition form factors and the correlation function parametrizing $B_c$ meson decay amplitudes are evaluated from the overlapping integrals of meson wave functions obtainable in different theoretical approaches. Some of them include the potential model approach \cite{Chang:1992pt}, the Bethe-Salpeture approach \cite{AbdEl-Hady:1999jux,Liu:1997hr}, relativistic constituent quark model on the light front \cite{Anisimov:1998xv,Wang:2008xt}, three-point sum rule of QCD, and non-relativistic QCD (NRQCD) \cite{Kiselev:2002vz}, relativistic quark model based on the quasi-potential approach \cite{Ebert:2003cn}, the Baur, Stech-Wirbel framework \cite{Dhir:2008hh}, perturbative QCD(PQCD) approach \cite{Wang:2012lrc,Rui:2016opu}, and the lattice QCD approach \cite{Lytle:2016ixw,Harrison:2020gvo}. Among the plethora, we have adopted relativistic independent quark (RIQ) model, that has been applied in wide-ranging hadronic sector describing the decay properties in the analysis of the magnetic dipole and electromagnetic transitions of $B_c$ and $B_c^*$ mesons in their ground as well as excited states\cite{Patnaik:2017cbl,Patnaik:2018sym}, the exclusive semileptonic $B_c$-mesons decays to the charmonium ground states in the vanishing \cite{Patnaik:2019jho} and non-vanishing \cite{Nayak:2021djn,Patnaik:2022moy,Patnaik:2023efe,Patnaik:2023ins} lepton mass limit. In view of the recent LHCb measurement \cite{LHCb:2020cyw} of the ratios of branching fraction in $B_s \to D_s l \nu_l$ to $B_s \to D_s^* l \nu_l$, we have also constructed a new ratio in $B_c$ decays to charm and charmonium states, which is given as:
\begin{align*}
    \mathcal{R}_{D/D^*} = \frac{\mathcal{B}\;(B_c \to D \mu \nu_{\mu})}{\mathcal{B}\;(B_c \to D^* \mu \nu_{\mu})} = 0.146, \\
    \mathcal{R}_{\eta_c/J/\psi} \frac{\mathcal{B}\;(B_c \to \eta_c \mu \nu_{\mu})}{\mathcal{B}\;(B_c \to J/\psi \mu \nu_{\mu})} = 0.140.
\end{align*}
Here the larger value for  
$\mathcal{R}_{D/D^*}$ than $ \mathcal{R}_{\eta_c/J/\psi}$ implies stronger relative preference for decaying to $D$ and $D^*$  than $\eta_c$ and $J/\psi$. This difference is attributed to the reduced phase space available in the latter system, resulting in a relatively higher value for $\mathcal{R}_{D/D^*}$. These ratios of branching fractions between pseudoscalar and vector states provides a window into the spin, angular momentum, and interaction dynamics of the $B_c$ meson, offering clues about the internal structure and interaction mechanisms.  These ratios of branching fractions serves as a powerful experimental tool as it enhances the ability to test theoretical models, reduces sensitivity to experimental and theoretical uncertainties, and provides a pathway to discovering new physics. Comparing this ratio to experimental data can provide evidence for or against different theoretical models of particle interactions, such as those involving the SM or ND-SM scenarios. 

We also predict new LFU observables in excited charm states of $B_c$ decays
\begin{align}
    \mathcal{R}_D(2S) = 0.656 \quad \mathcal{R}_D^*(2S) = 0.435\\
    \mathcal{R}_D(3S) = 0.464 \quad \mathcal{R}_D^*(3S) = 0.4
\end{align}
It is important to emphasize that our RIQ model does not utilize any adjustable free parameters. The potential parameters ($a$, $V_0$), quark masses ($m_q$), and the corresponding binding energies ($E_q$) were fixed based on hadron spectroscopy in the initial application of this model, primarily to replicate the hyperfine splittings of baryons and mesons. This fixed set of parameters has been consistently employed across a wide range of hadronic phenomena, as previously discussed. As a result, there are no free parameters within our model framework that could introduce uncertainties. The aim of calculating this new ratio \& LFU observables is to explore the applicability of the RIQ model and to conduct a qualitative analysis with respect to future enhanced collider data and improved theoretical models. Therefore, the main theoretical uncertainties in our predictions arise solely from the uncertainties in the input parameters: $V_{\rm cb} = 0.0408 \pm 0.0014$, $V_{\rm ub} = 0.00382 \pm 0.00020$. We have omitted these uncertainties in the $\mathcal{R}$ ratios, as the $V_{CKM}$ elements cancel out, leaving few significant sources of error in these observables.


Lattice QCD calculations are generally more accurate in this region because the daughter meson has small spatial momentum in the lattice frame (where the parent meson is usually taken to be at rest).The $B_c \to J/\psi$ form factor calculation that we describe here acts as a prototype for upcoming calculations of form factors for $B\rightarrow D^*$ and $B_s \rightarrow D_s^*$. $B_c \rightarrow J/\psi$ is a more attractive starting point for lattice QCD, however, because the mesons are 'gold-plated' (with tiny widths) and the valence quarks involved are all relatively heavy. This means that the valence quark propagators, from which appropriate correlation functions are constructed, are inexpensive to calculate. The correlation functions then have small statistical errors, even when the daughter has maximum spatial momentum. The absence of valence light quarks means that finite-volume effects are negligible and sensitivity to the u/d quark mass in the sea should be small. The main obstacle for the calculation of $B_c \to J/\psi$ form factors is that of the discretization effects associated with the heavy quarks. The c quark is handled very accurately in lattice QCD as long as improved discretization of the Dirac equation are used. All these anomalies can also be tested in the rare semileptonic decay of $B_c$ meson where the decay channels 
$B_c \to D^{(*)} l^+ l^-$ and $B_c \to D_s^{(*)} l^+ l^-$ can also prove to be important candidates for the search towards any ND-SM. These modes are yet to be identified precisely by the world wide experimental facilities as well as the lattice simulations. The obvious reasons behind this is very less branching fractions making it difficult to probe in presence of background. Also $B_c$ mesons are produced less frequently with compared to other bottom mesons and that is also the reason why the excited states are not explored well yet. Very recently, LHCb collaboration \cite{LHCb:2023lyb} also set the upper limit for the channel $B_c^+ \to D_s^+ \mu^+ \mu^-$ along with the fragmentation fractions of B meson with c and u quarks at 95\% confidence interval. Perturbative calculations for the hard associative production of two heavy pairs of $c \bar c$ and $b \bar b$ and a soft non-perturbative binding of non-relativistic quarks in the color-singlet state can be described in the framework of potential models. These two conditions result in the suppression of the $B_c$ yield of the order of $10^{-3}$ with respect to beauty hadrons

\subsection{Leptonic Decays}
Leptonic decay rates require decay constants to be determined. The $B_c$ meson leptonic decays, $B_c \to l \nu_l$ with l being e, $\mu$, or $\tau$, are heavy-quark- annihilation processes through axial-vector current, which are very important to the study of $B_c$ physics while have not been, but expected to be, measured. The primary leptonic decay mode of the $B_c$ meson involves the $\tau$ and $\nu_{\tau}$ particles. Nevertheless, detecting this decay experimentally is challenging due to the presence of hadronic backgrounds in the $\tau$ decays or the issue of missing energy. For experimental investigation of $B_c^{*+}\to l^+\nu_l$ decays, there should be at least more than $10^7$ $B_c^*$ events available. As of now, it is expected that more than $10^{12}$ Z bosons can be available at the future $e^+e^-$ colliders of CEPC \cite{dong2018derived} and FCC-ee \cite{abada2019fcc}. With BF ${\cal B} (Z\to b\bar{b}) = 12.03\pm0.21\%$ \cite{ParticleDataGroup:2024cfk} and fragmentation fraction $f (b\to B_c^*) \sim 6\times 10^{-4}$ \cite{boroun2016laplace, yang2019relativistic, zheng2019qcd}, there will more than $10^8$ $B_c^*$ events to search for $B_c^{*+}\to e^+\nu_e, \mu^+\nu_\mu, \tau^+\nu_\tau$ decays. In addition, the $B_c^*$ production cross sections at LHC are estimated to be about 100 nb for pp collisions at $\sqrt{s}$ = 13 TeV \cite{chen2018b} yielding more than $3\times 10^{10}$ $ B^*_c$ events corresponding to a dataset of 300 f$b^{-1}$ at LHCb. Hence, the $B_c^{*+}\to e^+\nu_e, \mu^+\nu_\mu, \tau^+\nu_\tau$ decays are expected to be carefully measured at LHCb experiments in the future.

Theoretically the decay rates of leptonic modes can be formulated as:
\begin{equation}
    \Gamma(B_c \to l \nu_l) = \frac{1}{8\pi} |V_{cb}|^2 G_F^2  M f^2_{B_c} m_l^2 \Biggl(1 - \frac{m_l^2}{M^2}\Biggr)^2
\end{equation}
where $V_{cb}$ denotes the CKM matrix element, M and $m_l$ stand for masses of $B_c$ meson and charged leptons,respectively;and $G_F$ is the Fermi coupling constant of weak interaction. Generally, the $B_c$ decay constant $f_{B_c}$ is defined through the transition matrix element of charged weak current, as
\begin{equation}
    \langle 0| b \gamma^{\mu}\gamma_5 c| {B_c(p)}\rangle
\end{equation}
which parameterizes the strong interaction effects and contains both perturbative and
non-perturbative contributions.

The available experimental data \cite{ParticleDataGroup:2024cfk}: $({\cal{R}}_\mu^\tau)^D=\frac{{\cal{B}}(D^+\to \tau^+\nu_\tau)}{{\cal{B}}(D^+\to \mu^+\nu_\mu)}=3.21\pm 0.73$,
$({\cal{R}}_\mu^\tau)^{D_s}=\frac{{\cal{B}}(D_s^+\to \tau^+\nu_\tau)}{{\cal{B}}(D_s^+\to \mu^+\nu_\mu)}=9.82\pm 0.40$ are consistent with SM expectations. Similar observables in $B_c$ decays, $(R_\mu^\tau)^{B_c}$ and $(R_\mu^\tau)^{B_c^*}$ have not yet to been measured. Therefore, prior to the experimental estimation, we have provided predictions for these observables in our QCD-inspired RIQ model framework, given as:
\begin{equation}
(\mathcal{R}_\mu^\tau)^{B_c} = \frac{\mathcal{B} (B_c \to \tau \nu_{\tau})}{\mathcal{B} (B_c \to \mu \nu_{\mu})} = 246.77 
\end{equation}
\begin{equation}
(\mathcal{R}_\mu^\tau)^{B_c^*} = \frac{\mathcal{B} (B_c^* \to \tau \nu_{\tau})}{\mathcal{B} (B_c^* \to \mu \nu_{\mu})} = 0.883
\end{equation}
Leptonic decays involve fewer QCD effects compared to semi-leptonic decays, where the final-state hadron introduces additional uncertainties. This reduces complications and makes LFU measurements in leptonic decays more robust for comparison with theoretical models. Therefore, LFU measurements in the leptonic decays of $B_c$ mesons offer an independent and complementary way to explore the same phenomena. If similar LFU violations are observed in $B_c \to  \tau \nu_{\tau}$ or $B_c \to \mu \nu_{\mu}$ it would strengthen the case for ND-SM physics. For more details on the predictions of $B_c$ decays constant, refer to \cite{Dash:2023ohf}. It has been shown in Ref.~\cite{Alonso:2016oyd} that the lifetime of the $B_c$ meson plays a crucial role in determining the Lorentz structures of scalar NP interaction. In SM, the lifetime of the $B_c$ meson ($\tau_{B_c}$) sets a limit on the yet unmeasured branching ratio $\mathcal{B}(B_c\to \tau\nu)$, which should not exceed the total decay width of the $B_c$ meson. This constraint has a significant effect on the couplings of scalar UP scenarios. The calculated value in SM is $\tau_{B_c}=0.52_{-0.12}^{+0.18}$ ps, obtained through operator product expansion~\cite{Chang:2000ac,Beneke:1996xe}. Importantly, this theoretical prediction is in good agreement with the experimental measurement of $\tau_{B_c}=0.507(9)$ ps~\cite{ParticleDataGroup:2022pth}. 

\subsection{Non-Leptonic Decays}
Non-leptonic two-body decays of $B_c$ mesons, while straightforward in terms of the weak decay of the b quark, are complicated by strong-interaction effects. Accurately computing these effects would greatly improve our ability to identify the source of CP violation in weak interactions, using the extensive data on these decays being gathered at the B factories. The two-body non-leptonic $B_c$ decays have been widely studied using various theoretical approaches and phenomenological models. The effective Hamiltonian describing the $B_c$ non-leptonic decays into meson states ($M_1, M_2$) is given by:
\begin{equation}
  \mathcal{A}(B_c \to M_1 M_2) = \frac{G_F}{\sqrt{2}} \lambda_c \sum_{i} {\cal C}_i(\mu) \langle M_1M_2|{\cal O}_i(\mu)|B_c\rangle_i
\end{equation}
\begin{equation}
    {\cal} H_{eff} = \frac{G_F}{\sqrt2} |V_{cb}| |V_{cq}|^\dag \sum_{i=1}^{6} {\cal} C_i {\cal} O_i
\end{equation}
In exclusive non-leptonic modes
usually involves the approximation of factorization \cite{Buras:1994ij} which, as expected, can be quite accurate for the $B_c$, since the quark-gluon sea is suppressed in the heavy quarkonium.. In the factorization hypothesis, non-leptonic meson decays can be categorized into three classes: class I, involving only color-favored tree diagrams (external W-emission diagrams) with a decay amplitude proportional to the QCD factor $a_1$; class II, involving only color-suppressed tree diagrams (internal W- emission) with a decay amplitude proportional to $a_2$; and class III, involving both external and internal W-emission contributions. Thus, the important parameters are the factors $a_1$ and $a_2$ in the non-leptonic weak Lagrangian, which depend on the normalization point suitable for the $B_c$ decays. The comprehensive predictions for non-leptonic decay modes of the $B_c$
meson within the RIQM framework are presented in~\cite{Dash:2023hjr}.
\section{CPV effects in \texorpdfstring{$B_c$}{} meson}
\label{sec_cp}
Since the identification of the CP-violating phenomenon in the decay of neutral kaons \cite{Christenson:1964fg}, numerous endeavors have been dedicated to investigating similar effects, particularly in beauty decays \cite{Fleischer:1997qs}. 
The CPV in the $B_c$ decays can be investigated in the same manner as made in the B decays. However, comparatively little focus has been directed towards studying these phenomena in $B_c$ meson decays \cite{Masetti:1992in}. Therefore it is particularly intriguing to explore the CPV effects in $B_c$ mesons using upcoming colliders.

In $B_c$ decays, various diagram types may contribute, including the color-favored tree diagram (external W emission diagram), the color-suppressed tree diagram (internal W emission diagram), the tree-annihilation diagram (W-annihilation diagram), the spacelike penguin diagram, and the timelike penguin diagram. Electroweak penguin diagrams contributions are minor compared to QCD penguin diagrams. In $B_c$ decays, CPV effects arise from contributions by color-favored tree diagrams, color-suppressed tree diagrams, and QCD penguin diagrams. Among these, tree diagram contributions are the most significant. The interference between tree diagrams and QCD penguin diagrams plays a much more critical role than the interference between the penguin diagrams themselves \cite{Liu:1997hr}.

In $B_c$ decays, only direct CP violation is possible, necessitating at least two contributing diagrams with different weak phases from the CKM matrix \cite{Cabibbo:1963yz} and distinct strong phases from final state interactions. However, the direct study of CP-violation in the $B_c$ decays is practically difficult because of low relative yield of $B_c$ with respect to ordinary B mesons: $\sigma_{B_c}/\sigma_{B} \sim 10^{-3}$. With QCD penguin diagrams contributing, hard strong phases can be roughly estimated perturbatively, and CP violation arises from interference among tree diagrams, QCD penguin diagrams, and within QCD penguin diagrams themselves. However, this estimation does not encompass potential soft strong phases from final state interaction. When one neglects soft strong phases, CPV in $B_c$ decays is restricted to processes with at least QCD penguin diagrams contributing. From an experimental perspective, the best $B_c$ decay modes that are promising for testing CP violation include $B_c \to J\psi D^*$ and $B_c \to \eta_c D^{(*)}$. The decay channels to exhibit CP asymmetries on the order of 1-2\% \cite{Masetti:1992in}. To observe these effects, approximately $10^7 - 10^8$ $B_c$ events would be needed, a quantity that should be feasible with the upcoming High-Luminosity LHC (HL-LHC).
\section{Future prospects \& potential developments}
\label{sec_future}
The calculations for exclusive $B_c$ decays will likely rely on lattice QCD, and new developments are needed to extend the calculations for the full ranges of $q^2$. Amidst this, we adopted QCD-inspired pehnomenological approach, which explicitly determines the dependence of form
factors for $B_c$ transition across the entire
kinematic range. This is achieved through overlap integrals of initial
and final meson wave functions, considering the confinement of constituent
quarks in the hadron core via an interaction potential with a suitable Lorentz
structure. This method represents a significant achievement in providing
phenomenological predictions and making our results valuable for comparison
with LQCD and other theoretical models. While, we do not claim high quantitative precision due to the absence of free parameters
in our calculations, we therefore achieve similar order of magnitude
with other models. By extending the RIQ model to different sectors, we
demonstrate its utility as an alternative phenomenological framework for
studying various hadronic phenomena and comparing results with other theoretical
efforts and experimental data. 
Ultimately, determining which theoretical approach is closest to the true,
QCD description will be resolved through experimentation. With upcoming
LHCb and Super-B experiments expected to provide high-statistics $B_c$
events, analyzing these decay modes within suitable phenomenological models
will be increasingly relevant.

Taking into account isospin violation and QED corrections presents a considerable difficulty for lattice QCD calculations. These have only begun to be discussed in certain circumstances; they must be covered in greater detail~\cite{Belle24:Lqcd}. Another possible improvement includes updating the $B_c$ form factors to a recently measured set that is more accurate. The new central values of the form factors corrections could potentially change the shape of the signals, but most importantly, the new form factors uncertainties will be smaller. Hence, considering that the theoretical uncertainty of the current $R_{J/\psi}$ analysis accounts for 18\% of the total uncertainty, updating the $B_c$ form factors with the latest theoretical calculations becomes crucial for refining the precision of the analysis. Similar prediction for new Ratios in excited charm and charmonium states will also open up new avenues to understand this heavy meson decays. The understanding of production rates in $B_c$ decays may present the biggest obstacle to precision physics. Since these decay modes are generic probes of numerous UP scenarios \cite{Cohen:1996vb} where the third generation is treated differently from the first and second, these decays are significant even if the current anomalies become less significant.

The $B_c$ meson's weak decay mechanism distinguishes it from charmonia and bottomonia, making it an excellent candidate for probing physics beyond the SM. Studies of its production and decay shed light on strong interactions in both perturbative and non-perturbative domains, with factorization theories like non-relativistic QCD providing valuable insights.

Apart from theoretical expectations, in principle, the two main general-purpose flavor factories are LHCb and Belle 2, which are scheduled to gather approximately 50 ${\rm fb}^{-1}$ of pp and 50 ${\rm fb}^{-1}$ of ee collision data by the early 2030s, respectively.  After Runs 3 and 4, LHCb intends to further upgrade the detector to meet the demands of the High Luminosity LHC. A luminosity around $10^{34}\text{cm}^{-2}\text{s}^{-1}$ will result in several dozen proton-proton collisions \cite{Koppenburg:2023ndc}. 
With the achievement of high statistics needed to extract the entire energy and angular distributions in $B_c$ decays, future high-luminosity LHCb \cite{LHCb:2018roe} and Belle II \cite{Belle-II:2018jsg} experiments and then it can be able to definitively address the $R(D^{(*)})$ anomalies that are currently being observed. The luminosity projections likelihood demonstrates that if the 2018 luminosity is increased by a factor of 3, the expected outcome is a straightforward decrease in the statistical component of the uncertainty by ${1}/{\sqrt 3}$ \cite{LHCb:2018roe}. This reduction could be achieved by incorporating the remaining data from Run 2, with a total integrated luminosity of 137 ${\rm fb}^{-1}$, and the dataset from Run 3, which has collected 67.37 ${\rm fb}^{-1}$ to date. 

Despite its intricate structure and considerable uncertainties making it less prominent, the journey of $B_c$ meson, in summary, is always important and exciting in view of probing small region of theory space.

\section*{Acknowledgements}
S.P. duly acknowledge the kind hospitality with support of PPC 2024, IIT Hyderabad, where this work is completed. S.P is grateful for substantial contribution from Profs. N. Barik, P. C. Dash, and S. Kar in the development of the RIQ model framework. S.P. acknowledge the NISER, Department of Atomic Energy, India for the financial support.




\bibliographystyle{utphys}
\bibliography{bibliography}






\end{document}